\documentclass[12pt,a4paper]{article}
\usepackage{amsfonts,latexsym}
\usepackage{amsmath,amssymb}
\usepackage{graphicx,color}
\usepackage{multirow}
\usepackage{mathdots}
\numberwithin{equation}{section} \oddsidemargin 0 mm \evensidemargin
0 mm \topmargin -10 mm \textheight 215 mm \textwidth 163 mm

\renewcommand{\thefootnote}{\fnsymbol{footnote}}
\newcommand{\nn}{\nonumber}

\begin{document}

\vspace{12mm}

\begin{center}
{{{\Large {\bf Gregory-Laflamme instability of BTZ black hole in new massive gravity}}}}\\[10mm]

{Taeyoon Moon\footnote{e-mail address: tymoon@sogang.ac.kr} and  Yun
Soo Myung\footnote{e-mail address: ysmyung@inje.ac.kr},
}\\[8mm]

{Institute of Basic Sciences and Department of Computer Simulation, Inje University, Gimhae 621-749, Korea}\\[0pt]

\end{center}
\vspace{2mm}

\begin{abstract}
We find the Gregory-Laflamme $s$-mode instability of the
non-rotating BTZ black hole in new  massive gravity. This
instability shows that the BTZ black hole could not exist as a
stable static solution to the new massive gravity. For non-rotating
BTZ black string in four dimensions, however, it is demonstrated
that the BTZ black string can be stable against the metric
perturbation.
\end{abstract}
\vspace{5mm}


\vspace{1.5cm}

\hspace{11.5cm}{Typeset Using \LaTeX}
\newpage
\renewcommand{\thefootnote}{\arabic{footnote}}
\setcounter{footnote}{0}

\section{Introduction}
The dRGT gravity~\cite{deRham:2010ik,deRham:2010kj,Hassan:2011hr} is
considered as a promising massive gravity model which yields
Einstein gravity in the massless limit. Recently, it was shown that
the stability of the Schwarzschild black hole in the four
dimensional dRGT gravity could be  determined by the
Gregory-Laflamme (GL) instability
\cite{Gregory:1993vy,Gregory:1994bj} of a five-dimensional black
string.  The small Schwarzschild black hole with mass $M_{\rm S}$ in
the dRGT gravity and its bi-gravity
extension~\cite{Babichev:2013una,Brito:2013wya}, and fourth-order
gravity~\cite{Myung:2013doa} is unstable against  metric and Ricci
tensor perturbations for $m' \le \frac{{\cal O}(1)}{M_{\rm S}}$ and
$m_2 \le \frac{1}{2M_{\rm S}}$, respectively. These results may
indicate that static black holes in massive gravity  do  not exist.

 Interestingly, it
turned out that in a massive theory of the Einstein-Weyl gravity,
the linearized Einstein tensor perturbations exhibit unstable modes
of the Schwarzschild-AdS black hole featuring the GL instability of
five-dimensional AdS black string, in contrast to the stable
Schwarzschild-AdS black hole in the Einstein
gravity~\cite{Myung:2013bow}. The linearized Ricci tensor
perturbations were employed to exhibit unstable modes of the
Schwarzschild-Tangherlini (higher dimensional Schwarzschild) black
hole in higher-dimensional fourth order gravity which features the
GL instability of higher dimensional black
strings~\cite{Myung:2013cna}, in comparison with  the stable
Schwarzschild-Tangherlini  black holes in higher-dimensional
Einstein gravity. These imply that the GL instability of the  black
holes in the massive gravity originates from the massiveness, but
not a nature of  the fourth-order gravity  giving ghost states.
Also, one could avoid  the ghost problem arising from the metric
perturbations in the fourth-order gravity when using the linearized
Einstein and Ricci tensors because their linearized equations become
the second-order tensor equations.

On the other hand, it was shown that the four-dimensional  BTZ black
string in Einstein gravity is stable against  metric perturbations
regardless of the horizon size, which is also supported by a
thermodynamic argument of Gubser-Mitra
conjecture~\cite{Kang:2002hx,Kang:2006zh}. Later on, however,  it
was argued that the BTZ black string is not always stable against
metric perturbations~\cite{Liu:2008ds}.  In the
literatures~\cite{Kang:2002hx,Kang:2006zh}, there exists a threshold
value for $\mu^2>0$ (we use a different notation $\mu^2$ to avoid a
confusion $m^2$ here) which is related to the compactification of
the extra dimension of the tensor perturbation. It was shown
in~\cite{Liu:2008ds} that for $\mu^2 \ge 3/\ell^2$ with $\ell$
AdS$_3$ curvature radius, the BTZ black string is stable against
$s$-mode metric perturbation, while for $\mu^2<3/\ell^2$ it is
unstable. Therefore, it seems to be necessary to point out which one
is correct.

The new massive gravity  has been introduced as a fourth-order
gravity with a healthy massive spin-2 mode and a massless spin-2
ghost mode, which is pure gauge only in three
dimensions~\cite{Bergshoeff:2009hq}. This parity-even gravity
describes two  modes of helicity $+2$ and $-2$ [2 degrees of freedom
(DOF)] of a massive graviton, but it has a drawback of serving as a
unitary model of massive gravity only in three
dimensions~\cite{Bergshoeff:2013vra}.
 For $m^2> 1/2\ell^2$ with $m^2$ the mass of graviton,  the
three-dimensional BTZ black hole in the new massive gravity is shown
to be stable against $s$-mode metric
perturbation~\cite{Myung:2011bn} when using the positivity of the
potential outside the horizon.  To this direction, it was recently
reported  that the stability of the BTZ black hole was mainly
determined by the asymptotes of black hole spacetime: the condition
of the $s$-mode stability is consistent with the generalized
Breitenlohner-Freedman (BF) bound ($m^2\ge -1/2\ell^2$) for metric
perturbations on asymptotically AdS$_3$
spacetime~\cite{Moon:2013jna}. This result
 may  imply   that the stability condition is extended simply from
$m^2> 1/2\ell^2$ to $m^2\ge -1/2\ell^2$ if $m^2$ is allowed to be  a
negative quantity. However, one expects that two different type of
instabilities appears for the BTZ black hole in new massive gravity:
one is from the BF bound based on the tensor propagation on
asymptotically AdS$_3$ spacetime, while the other is the GL
instability of a massive graviton propagating on the BTZ black hole
spacetime. This is similar to two instabilities of AdS black holes
to trigger a holographic superconductor phase within the AdS/CFT
correspondence~\cite{Hartnoll:2009sz}. In these models, the AdS$_d$
black hole becomes unstable to form non-trivial fields outside its
horizon when being close to extremality whose near-horizon geometry
is AdS$_2\times M_{d-2}$. For a massive scalar
 with mass $m^2$ between
$-(d-1)^2/4\ell^2$ and $-1/4\ell^2$, two AdS spacetimes  are
unstable~\cite{Hartmann:2013nla}.

 Hence, it suggests strongly that
the stability of BTZ black hole should  be revisited in new massive
gravity  by observing  the GL instability of four-dimensional black
string. We will show that the instability of a massive graviton
persists even in three-dimensional BTZ black hole. This establishes
that the instability of the black holes in the $D\ge 3$-dimensional
massive gravity  originates from the massiveness, but not a nature
of  the fourth-order gravity giving ghost states. As a byproduct, we
will show that the four-dimensional BTZ black string is stable
against the $s$-mode metric perturbation because  its mass squared
is positive ($\mu^2>0$).
\section{Linearized perturbation equation}
We start with the three-dimensional fourth order gravity defined as
\begin{eqnarray}\label{3dfog}
S_{\rm fog}=\int
d^3x\sqrt{-g}\Bigg[\frac{R-2\lambda_S}{\kappa}+\alpha R^2 +\beta
R_{\mu\nu}R^{\mu\nu}\Bigg]
\end{eqnarray}
with $\kappa$ the three-dimensional gravitational coupling constant.
 From the action (\ref{3dfog}),  the
Einstein equation is derived to be
\begin{equation} \label{equa1}
\frac{1}{\kappa}\Big(R_{\mu\nu}-\frac{1}{2} Rg_{\mu\nu}+\lambda_S
g_{\mu\nu}\Big)+E_{\mu\nu}=0,
\end{equation}
where $E_{\mu\nu}$ takes the form
\begin{eqnarray} \label{equa2}
E_{\mu\nu}&=& 2\beta
\Big(R_{\mu\rho\nu\sigma}R^{\rho\sigma}-\frac{1}{4}
R^{\rho\sigma}R_{\rho\sigma}g_{\mu\nu}\Big)+2\alpha
R\Big(R_{\mu\nu}-\frac{1}{4} Rg_{\mu\nu}\Big) \nonumber \\
&+&
\beta\Big(\nabla^2R_{\mu\nu}+\frac{1}{2}\nabla^2Rg_{\mu\nu}-\nabla_\mu\nabla_\nu
R\Big) +2\alpha\Big(g_{\mu\nu} \nabla^2R-\nabla_\mu \nabla_\nu
R\Big).
\end{eqnarray}

For $\Lambda=\lambda_S+2\kappa(3\alpha+\beta)\Lambda^2$ with
$\bar{R}_{\mu\nu}=2\Lambda \bar{g}_{\mu\nu}$,  the non-rotating BTZ
black hole solution is given by
\begin{equation} \label{btz}
  ds^2_{\rm BTZ}=\bar{g}_{\mu\nu}dx^\mu dx^\nu=-V(r)dt^2
   +\frac{dr^2}{V(r)}+r^2d\phi^2,
\end{equation}
where the metric function takes the form  \begin{equation}
V(r)=-{\cal M}+\frac{r^2}{\ell^2}
\end{equation}
with $\ell^2=-1/\Lambda$ and ${\cal M}$  the ADM mass. From the
condition of $V(r_+)=0$, the horizon is located at $r=r_+$.

For a perturbation around the BTZ black hole
\begin{eqnarray}\label{pert}
g_{\mu\nu}=\bar{g}_{\mu\nu}+h_{\mu\nu},
\end{eqnarray}
the linearized Einstein tensor, Ricci tensor, and Ricci scalar are
given by
\begin{eqnarray}
\delta G_{\mu\nu}(h)&=&\delta R_{\mu\nu}-\frac{1}{2} \bar{g}_{\mu\nu} \delta R-2\Lambda h_{\mu\nu}, \\
 \delta
R_{\mu\nu}(h)&=&\frac{1}{2}\Big(\bar{\nabla}^{\rho}\bar{\nabla}_{\mu}h_{\rho\nu}
+\bar{\nabla}^{\rho}\bar{\nabla}_{\nu}h_{\rho\mu}-\bar{\nabla}^2h_{\mu\nu}
-\bar{\nabla}_{\mu}\bar{\nabla}_{\nu}h\Big),\label{lrmunu}\\
\delta
R(h)&=&\bar{\nabla}_{\alpha}\bar{\nabla}_{\beta}h^{\alpha\beta}-\bar{\nabla}^2h-2\Lambda
h\label{lr},
\end{eqnarray}
where the Einstein tensor
$G_{\mu\nu}=R_{\mu\nu}-Rg_{\mu\nu}/2+\Lambda g_{\mu\nu}$. With the
help of the above quantities, the linearized Einstein equation can
be written as
\begin{eqnarray}
&&\Big[\frac{1}{\kappa}+(12\alpha+2\beta)\Lambda\Big]\delta
G_{\mu\nu} \nonumber
\\
&&+(2\alpha+\beta)\Big[\bar{g}_{\mu\nu}\bar{\nabla}^2-\bar{\nabla}_{\mu}
\bar{\nabla}_{\nu}+2\Lambda \bar{g}_{\mu\nu} \Big] \delta R
+\beta\Big(\bar{\nabla}^2\delta G_{\mu\nu}-\Lambda \bar{g}_{\mu\nu}
\delta R\Big)=0. \label{lineq}
\end{eqnarray}
Taking the trace of (\ref{lineq}) provides the linearized Ricci
scalar equation \begin{equation}
\Big[(8\alpha+3\beta)\bar{\nabla}^2-\Big\{\frac{1}{\kappa}-4(3\alpha+\beta)\Lambda\Big\}\Big]
\delta R=0,
\end{equation}
which implies that for $8\alpha+3\beta=0$ (the new massive gravity),
the d'Alembertian operator is removed. In the new massive
gravity~\cite{Bergshoeff:2009hq}
\begin{eqnarray}\label{snmg}
S_{\rm NMG}=\int
d^3x\sqrt{-g}\Bigg[\frac{R-2\lambda_S}{\kappa}+\beta\left(
R_{\mu\nu}R^{\mu\nu}-\frac{3}{8}R^2\right)\Bigg],
\end{eqnarray}
 $\delta R$ is
constrained to vanish \begin{equation} \delta R=0 \end{equation}
provided that $\kappa\beta\Lambda\not=-2$. This explains why we
choose the new massive gravity. Plugging $\delta R=0$ and
$\alpha=-3\beta/8$ into (\ref{lineq}) leads to the linearized
Einstein tensor equation
\begin{equation} \label{linein}
\Big(\bar{\nabla}^2-2\Lambda -M^2\Big) \delta G_{\mu\nu} =0,
\end{equation}
where the mass squared is given by
\begin{equation} \label{mass-sq}
M^2~=-\frac{1}{\kappa\beta}+\frac{\Lambda}{2}\equiv
m^2-\frac{1}{2\ell^2}.
\end{equation}
Choosing the transverse-traceless gauge
\begin{equation} \label{tt}
\bar{\nabla}^\mu h_{\mu\nu}=0,~~h=0,
\end{equation}
Eq. (\ref{linein}) leads to the fourth-order equation for the metric
perturbation $h_{\mu\nu}$
\begin{equation}
\Big(\bar{\nabla}^2-2\Lambda\Big) \Big(\bar{\nabla}^2-2\Lambda
-M^2\Big) h_{\mu\nu} =0, \label{linh}
\end{equation}
which might imply  the two second-order linearized equations
\begin{eqnarray}\label{nmgmeq1}
&&\Big[\bar{\nabla}^2-2\Lambda\Big]h_{\mu\nu}=0,
\\ \label{nmgmeq2}
&&\Big[\bar{\nabla}^2-2\Lambda -M^2\Big]h_{\mu\nu}=0
\end{eqnarray}
off critical point ($M^2\not=0,~m^2\neq1/2\ell^2$). Even though  Eq.
(\ref{nmgmeq2})  may describe 2 DOF for a massive graviton in three
dimensions, it might not be a correct equation because the `$-$'
sign disappears when splitting (\ref{linh}) into (\ref{nmgmeq1}) and
(\ref{nmgmeq2}). In order to see this ghost problem explicitly, we
consider the three-dimensional flat spacetime. In the case of
$\Lambda\to 0~(\lambda_S \to0)$, the tree-level scattering amplitude
$A$ between two conserved sources $T'_{\mu\nu}$ and $T_{\mu\nu}$ is
given by~\cite{Gullu:2009vy}
\begin{equation}
4A=-\frac{2}{\beta} T'_{\mu\nu}\frac{1}{p^2(p^2+m^2)}T_{\mu\nu} +
\frac{1}{\beta}T'\frac{1}{p^2(p^2+m^2)}T-\kappa T'\frac{1}{p^2}T
\end{equation}
with $p^2=-\partial^2$. After partial fractions, this leads to
\begin{equation}
4A=2\kappa
T'_{\mu\nu}\Big(\frac{1}{p^2}-\frac{1}{p^2+m^2}\Big)T_{\mu\nu}
-\kappa T'\Big(\frac{2}{p^2}-\frac{1}{p^2+m^2}\Big)T.
\end{equation}
In order not to have a tachyon, we have to choose $\kappa\beta <0
~(m^2>0)$ with $\beta>0$. Also, we require $\kappa<0$ to have  a
healthy massive spin-2 without negative norm states (ghosts) and a
massless spin-2 ghost. This is why one chooses a negative $\kappa$
for a positive $\beta$ in 3D flat spacetime. However, we might miss
the ghost problem when splitting (\ref{linh}) into (\ref{nmgmeq1})
and (\ref{nmgmeq2}) without imposing the sign correction.  Hence, it
would be better to use the second-order equation (\ref{linein}) for
the linearized Einstein tensor if it could describe 2 DOF. For this
purpose, we have  two constraints
\begin{equation} \label{gtt}
\delta G=-\delta R=0,~~~ \bar{\nabla}^\mu \delta G_{\mu\nu}=0,
\end{equation}
where the last one  comes from contracting  the linearized Bianchi
identity $\bar{\nabla}_{[\tau} \delta R_{\mu\nu]\rho\sigma}=0$ with
$\bar{g}^{\tau\rho}\bar{g}^{\mu\sigma}$. Then, we have 2 DOF because
$6-1-3=2$.  Hence, we note that Eq. (\ref{linein}) is interpreted as
a  boosted-up version of Eq.
(\ref{nmgmeq2})~\cite{Bergshoeff:2013vra}.

In  3D flat spacetime, the mass squared  should be positive
($m^2>0$) because of the non-tachyonic condition but it is allowed
to be negative in asymptotically AdS$_3$ spacetime. Accordingly, the
BF bound for a tensor field could be read off from Eqs.
(\ref{linein}) and (\ref{nmgmeq2}) as
\begin{equation} \label{bf-bound}
M^2\ge -\frac{1}{\ell^2} \to m^2\ge -\frac{1}{2\ell^2}.
\end{equation}
However, in the case of 4D BTZ black string, one requires $\mu^2>0$
because it comes from the compactification of an extra
dimension~\cite{Liu:2008ds}. Hence, the BF bound does not require a
further condition for  the stability  of the 4D BTZ black string.

\section{ Gregory-Laflamme $s$-mode instability}

In order to investigate the GL instability of the massive graviton
propagating on the BTZ black hole spacetime, we first start with
(\ref{nmgmeq2}) for convenience, because (\ref{linein}) and
(\ref{nmgmeq2}) are the same equation. For $s(k=0)$-mode analysis, a
metric perturbation takes the form with four components
$H_{tt},~H_{tr},~H_{rr},$ and $H_3$ as
\begin{eqnarray}
h_{\mu\nu}(t,\phi,r)= e^{\Omega t}e^{ik\phi}|_{k=0}\left(
\begin{array}{ccc}
H_{tt}(r)  & H_{tr}(r) & 0  \cr H_{tr}(r) & H_{rr}(r) & 0 \cr 0 & 0
& H_3(r)
\end{array}
\right)\,. \label{type2}
\end{eqnarray}
Substituting Eq.(\ref{type2}) into Eq.(\ref{nmgmeq2}) and after a
tedious manipulation, we obtain the second-order equation for a
single physical field $H_{tr}(r)$ \cite{Myung:2011bn} as
\begin{eqnarray}\label{Htreq}
&&\left\{(m^2-1/2\ell^2)(r^2/\ell^2-{\cal M})+r^2/\ell^4-2{\cal
M}/\ell^2+\Omega^2\right\}H_{tr}''+\left\{ \frac{7r^2/\ell^2-{\cal
M}}{r(r^2/\ell^2-{\cal
M})}\Omega^2\right.\nonumber\\
&&\left.+\frac{5r^2/\ell^2-{\cal M}}{r}(m^2-1/2\ell^2)
+\frac{5r^4/\ell^4-13{\cal M}r^2/\ell^2+2{\cal M}^2}{\ell^2
r(r^2/\ell^2-{\cal M})}\right\}H_{tr}'+
\left\{\frac{6r^4/\ell^4-{\cal M}^2}{r^2(r^2/\ell^2-{\cal
M})^2}\Omega^2
\right.\nonumber\\
&&\left.+\frac{2r^4/\ell^4-2{\cal M}r^2/\ell^2-{\cal
M}^2}{r^2(r^2/\ell^2-{\cal M})}(m^2-1/2\ell^2)
+\frac{(3r^2/\ell^2-2{\cal M})(r^4/\ell^4-4{\cal M}r^2/\ell^2-{\cal
M}^2)}{\ell^2r^2(r^2/\ell^2-{\cal M})^2}
\right.\nonumber\\
&&\left.\hspace{3em}-\left(m^2-1/2\ell^2+\frac{\Omega^2}{r^2/\ell^2-{\cal
M}}\right)^2 \right\}H_{tr}=0
\end{eqnarray}
with ${\cal M}=r_+^2/\ell^2$.  This implies that the $s$-mode
perturbation is described by a single field $H_{tr}$, even though we
were starting with four components and the massive graviton has two
DOF for $k\not=0$.

It turns out that the  second-order equation (\ref{Htreq}) can be
reduced to two first-order equations  with a constraint when  using
the perturbation equation (\ref{nmgmeq2}) together with  the TT
gauge condition (\ref{tt})~\cite{Gregory:1994bj}. The  two coupled
first-order equations are given by
\begin{eqnarray}
\hspace*{-2em}H'&=&\frac{{\cal
M}-3r^2/\ell^2}{rV}H-\frac{\Omega}{2V}(H_++H_-),
\label{Hd}\\
&&\nn\\
\hspace*{-2em}H_{-}^{\prime}&=&\frac{\Omega}{{\cal
M}}H+\Big[\frac{1}{2r}-\frac{(2m^2+1/\ell^2)r}{4{\cal M}}\Big]H_++
\Big[-\frac{3}{2r}+\frac{(2m^2+1/\ell^2)r}{4{\cal
M}}+\frac{r\Omega^2}{{\cal M}V}\Big] H_-.\label{Hmd}
\end{eqnarray}
A constraint equation can be written as
\begin{eqnarray}
&&r\Omega\Big[-5{\cal M}/\ell^2+r^2/\ell^4+2m^2V+4\Omega^2\Big]H_-
-rV(2m^2+1/\ell^2)\Omega H_+\nn\\
&&\hspace*{15em}+2V(2m^2{\cal M}-{\cal
M}/\ell^2+2\Omega^2)H~=~0,\label{cons}
\end{eqnarray}
 where
\begin{eqnarray}
H\equiv -H_{tr},~~~H_\pm\equiv\frac{H_{tt}}{V(r)}\pm V(r)H_{rr}.
\end{eqnarray}
At infinity of $r\to\infty$, asymptotic solutions to Eqs.(\ref{Hd})
and (\ref{Hmd}) are
\begin{eqnarray}
H^{(\infty)}&=&C^{(\infty)}_{1}r^{-2-\sqrt{m^2\ell^2+1/2}}
+C^{(\infty)}_{2}r^{-2+\sqrt{m^2\ell^2+1/2}},\nn\\
&&\nn\\
H_-^{(\infty)}&=&\tilde{C}^{(\infty)}_{1}r^{-1-\sqrt{m^2\ell^2+1/2}}
+\tilde{C}^{(\infty)}_{2}r^{-1+\sqrt{m^2\ell^2+1/2}},
\end{eqnarray}
where $\tilde{C}^{(\infty)}_{1,2}$ are
\begin{eqnarray}
\tilde{C}^{(\infty)}_{1}~=~\frac{2m^2-1/\ell^2}
{\Omega\Big(2+\sqrt{4m^2\ell^2+2}\Big)}{C}^{(\infty)}_{1}
,~~~~~\tilde{C}^{(\infty)}_{2}~=~
\frac{2m^2-1/\ell^2}{\Omega\Big(2-\sqrt{4m^2\ell^2+2}\Big)}{C}^{(\infty)}_{2}.
\end{eqnarray}
 At the horizon of
$r_+=\ell\sqrt{{\cal M}}$,  their solutions are given by
\begin{eqnarray}
H^{(r_+)}&=&C^{(r_+)}_{1}(r-r_+)^{-1-\Omega\ell/(2\sqrt{ {\cal M}})}
+C^{(r_+)}_{2}(r-r_+)^{-1+\Omega\ell/(2\sqrt{{\cal
M}})},\label{Ha}\\
&&\nn\\
H_-^{(r_+)}&=&\tilde{C}^{(r_+)}_{1}(r-r_+)^{-\Omega\ell/(2\sqrt{
{\cal M}})} +\tilde{C}^{(r_+)}_{2}(r-r_+)^{\Omega\ell/(2\sqrt{{\cal
M}})},\label{Hma}
\end{eqnarray}
where $\tilde{C}^{(r_+)}_{1,2}$ take the forms
\begin{eqnarray}
\tilde{C}^{(r_+)}_{1}&=&\frac{{\cal
M}(-2m^2\ell^2+1)-2\Omega^2\ell^2+\Omega\ell(2m^2\ell^2+1)\sqrt{{\cal
M}}}{\Omega\ell({\cal M}-\Omega^2\ell^2)}~{C}^{(r_+)}_{1}
,\nn\\
\tilde{C}^{(r_+)}_{2}&=&\frac{{\cal
M}(-2m^2\ell^2+1)-2\Omega^2\ell^2-\Omega\ell(2m^2\ell^2+1)\sqrt{{\cal
M}}}{\Omega\ell({\cal M}-\Omega^2\ell^2)}~{C}^{(r_+)}_{2}.
\end{eqnarray}
Imposing two boundary conditions of  the regular solutions at
infinity and horizon  correspond to choosing ${C}^{(\infty)}_{2}=0$
and ${C}^{(r_+)}_{1}=0$, respectively.

Eliminating $H_+$ in Eqs. (\ref{Hd}) and (\ref{Hmd}) by using the
constraint (\ref{cons}) leads to the two  coupled equations with $H$
and $H_-$ only. For fixed $m$ and various values of $\Omega$, we
solve these equations numerically\footnote{Solving the first-order
equations (\ref{Hd}) and (\ref{Hmd}) numerically, we begin it by
considering  asymptotic solutions (\ref{Ha}) and (\ref{Hma}) at the
horizon. For given $m$, we find the consistent values of $\Omega$ to
have the asymptotic behavior of $H^{(\infty)}\sim
r^{-2-\sqrt{m^2\ell^2+1/2}}$. For a complete analysis, we have
already checked the results given in Fig.3 of the literature
\cite{Gregory:1994bj}.} which yields permitted values of $\Omega$ as
a function of $m^2$. As a result, this shows clearly that there
exist unstable modes (see Fig.1).
\begin{figure*}[t!]
   \centering
   \includegraphics{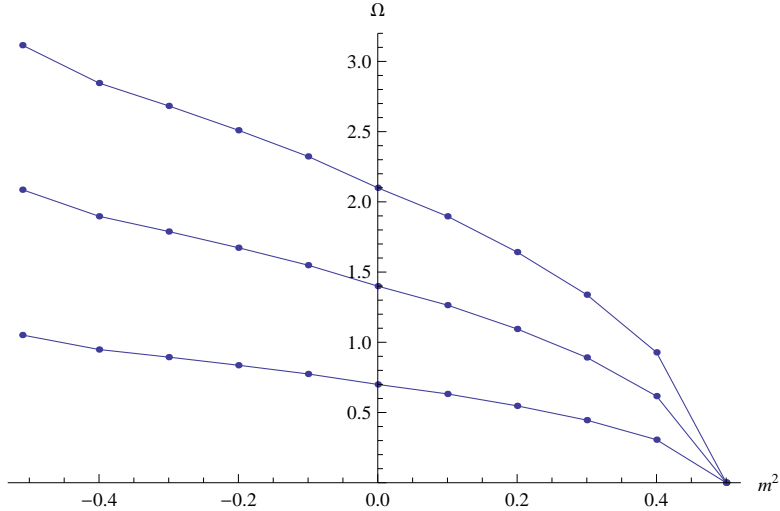}
\caption{$\Omega$ graphs are depicted as function of $m^2$ for three
different horizon radii of $r_+=1,~2,~3$ from bottom to top with
$\ell=1$. The data range for $m^2$ is between $-1/2$ and $1/2$.}
\end{figure*}

Comparing with higher dimensional black
strings~\cite{Gregory:1993vy,Gregory:1994bj}, we would like to
mention a couple of key points observed from  Fig. 1. Firstly we can
fix the AdS$_{3}$ curvature radius ($\ell=1$) by taking into account
the scaling symmetry given in Eqs. (\ref{Hd})-(\ref{cons}) as
\begin{eqnarray}
r\to\alpha r,~~~m\to
m/\alpha,~~~\Omega\to\Omega/\alpha,~~~\ell^2\to\alpha^2\ell^2
\end{eqnarray}
with an arbitrary constant $\alpha$. The second point is that the
threshold mass  for $r_+=1,2,3$ is given
 near $m^2\approx 0.5$, which implies
that GL instability exists for
\begin{equation}
m^2<\frac{1}{2\ell^2}
\end{equation}
when recovering the AdS$_3$ curvature radius $\ell$. This means that
the BTZ black hole is unstable against the $s$-mode metric
perturbation regardless of the horizon size.  On the other hand, the
threshold mass of $m^2=1/2\ell^2$ can be read off approximately by
taking the limit of $\Omega\to0$ in the constraint equation
(\ref{cons}). Therefore, the stability condition is given by
\begin{equation} \label{st-con}
m^2 > \frac{1}{2\ell^2}
\end{equation}
which is exactly the same  condition obtained from the positivity of
the potential ($V_\Psi>0$)~\cite{Myung:2011bn}. The potential
appears in the Schr\"odinger equation
\begin{equation}
\frac{d^2\Psi}{dr_*^2}+[\omega^2-V_\Psi]\Psi=0,~~\omega=i\Omega
\end{equation}
which was derived from the second-order equation (\ref{Htreq}) by
introducing a new field $\Psi$ defined by $\Psi=H_{tr}/f(r)$.
Combining it with the BF bound (\ref{bf-bound}) for stability
condition of tensor field at asymptotically AdS$_3$ spacetime, we
find that the instability of the BTZ black hole in new massive
gravity is extended to
\begin{equation} \label{int-con}
-\frac{1}{2\ell^2} \le m^2 <\frac{1}{2\ell^2}.
\end{equation}
Up to now, we have made our instability analysis with
(\ref{nmgmeq2}) for the metric tensor. Since two equations
(\ref{linein}) and (\ref{nmgmeq2}) take the same form when replacing
$\delta G_{\mu\nu}$ by $h_{\mu\nu}$ and they describe 2 DOF with
(\ref{tt}) and (\ref{gtt}), the instability analysis for $
h_{\mu\nu}$ persists in that of $\delta G_{\mu\nu}$. One additional
advantage when using (\ref{linein}) is to avoid the ghost problem.

Finally, considering $M^2=m^2-1/2\ell^2$ (\ref{mass-sq}), we rewrite
the stability condition (\ref{st-con}) as
\begin{equation}
M^2>0
\end{equation}
which is exactly the same condition  for the 4D black string with
positive mass squared $\mu^2>0$. This implies that the 4D black
string is stable under the $s$-mode metric perturbation regardless
of the horizon size~\cite{Kang:2006zh}.

\section{Discussions}

We have shown that the new massive gravity which is known to  be a
unitary gravity model in three dimensions, could not accommodate the
BTZ black hole by observing the GL instability for the mass $m^2$
between $-1/2\ell^2 $ and $ 1/2\ell^2$. The GL instability has
nothing to do with the ghost issue arising from the fourth-order
gravity of the new massive gravity because we have used two
second-order equations (\ref{nmgmeq2}) and (\ref{linein}) for
$h_{\mu\nu}$ and $\delta G_{\mu\nu}$, respectively. Also, this
instability could not be explained in terms of the Gubser-Mitra
conjecture because the heat capacity of the BTZ black hole is always
positive. This instability arises from the massiveness
($M^2=m^2-1/2\ell^2\not=0$) of the new massive gravity. In the
massless case of $M^2=0$, one has a massless graviton propagating on
the BTZ black hole spacetime which is  gauge artefact in three
dimensions. In this case, the stability issue of the BTZ black hole
is meaningless.

 This
establishes that the instability of the static black holes in the
$D\ge 3$-dimensional massive gravity originates from the
massiveness, but not a nature of the fourth-order gravity giving
ghost states.

Also, the unstable condition of (\ref{int-con}) explains that
 choosing the graviton  mass above the three-dimensional BF bound
 but below the GL instability bound has made the BTZ black hole
 unstable to a formation of massive graviton which is regular at
 horizon and infinity. This is similar to  two instabilities of AdS
black holes to realize a holographic superconductor phase in the
bulk within the AdS/CFT
correspondence~\cite{Hartnoll:2009sz,Hartmann:2013nla}.

Finally, we would like to mention that the 4D black string is stable
under the $s$-mode metric perturbation in Einstein
gravity~\cite{Kang:2006zh}, while the BTZ black hole is unstable in
the new massive gravity. This shows a newly interesting feature of
low dimensional black string and black hole when comparing with
higher dimensional black string and black holes.

\section*{Acknowledgement}

  T.M. would like to thank Dr. Miok Park for useful discussions.
  This work was  supported
by the National Research Foundation of Korea (NRF) grant funded by
the Korea government (MEST) (No.2012-R1A1A2A10040499). Y.M. was
supported partly by the National Research Foundation of Korea (NRF)
grant funded by the Korea government (MEST) through the Center for
Quantum Spacetime (CQUeST) of Sogang University with grant number
2005-0049409.


\newpage

\begin{thebibliography}{99}

\bibitem{deRham:2010ik}
  C.~de Rham and G.~Gabadadze,
   Phys.\ Rev.\ D {\bf 82}, 044020 (2010)  [arXiv:1007.0443
   [hep-th]].

\bibitem{deRham:2010kj}
  C.~de Rham, G.~Gabadadze and A.~J.~Tolley,
  Phys.\ Rev.\ Lett.\  {\bf 106}, 231101 (2011)  [arXiv:1011.1232
  [hep-th]].

\bibitem{Hassan:2011hr}
  S.~F.~Hassan and R.~A.~Rosen,
  Phys.\ Rev.\ Lett.\  {\bf 108}, 041101 (2012)  [arXiv:1106.3344 [hep-th]].


\bibitem{Gregory:1993vy}
  R.~Gregory and R.~Laflamme,
  Phys.\ Rev.\ Lett.\  {\bf 70}, 2837 (1993)
  [hep-th/9301052].


\bibitem{Gregory:1994bj}
  R.~Gregory and R.~Laflamme,
  Nucl.\ Phys.\ B {\bf 428} (1994) 399
  [hep-th/9404071].


\bibitem{Babichev:2013una}
  E.~Babichev and A.~Fabbri,
   Class.\ Quant.\ Grav.\  {\bf 30}, 152001 (2013)  [arXiv:1304.5992 [gr-qc]].

\bibitem{Brito:2013wya}
  R.~Brito, V.~Cardoso and P.~Pani,
  Phys.\  Rev.\ D {\bf 88}, 023514 (2013)  [arXiv:1304.6725 [gr-qc]].


\bibitem{Myung:2013doa}
  Y.~S.~Myung,
 Phys.\  Rev.\ D {\bf 88}, 024039 (2013)  [arXiv:1306.3725 [gr-qc]].

\bibitem{Myung:2013bow}
  Y.~S.~Myung,
  arXiv:1308.1455 [gr-qc].

\bibitem{Myung:2013cna}
  Y.~S.~Myung,
  Phys.\  Rev.\ D {\bf 88}, 084006 (2013) [arXiv:1308.3907 [gr-qc]].

\bibitem{Kang:2002hx}
  G.~Kang,
   hep-th/0202147.

\bibitem{Kang:2006zh}
  G.~Kang and Y.~O.~Lee,
   AIP Conf.\ Proc.\  {\bf 805}, 358 (2006).

\bibitem{Liu:2008ds}
  L.~-h.~Liu and B.~Wang,
   Phys.\ Rev.\ D {\bf 78}, 064001 (2008)  [arXiv:0803.0455 [hep-th]].

\bibitem{Bergshoeff:2009hq}
  E.~A.~Bergshoeff, O.~Hohm and P.~K.~Townsend,
  Phys.\ Rev.\ Lett.\  {\bf 102}, 201301 (2009)
  [arXiv:0901.1766 [hep-th]].

\bibitem{Bergshoeff:2013vra}
  E.~Bergshoeff, M.~Kovacevic, L.~Parra and T.~Zojer,
  PoS Corfu {\bf 2012}, 053 (2013).


\bibitem{Myung:2011bn}
  Y.~S.~Myung, Y.~-W.~Kim, T.~Moon and Y.~-J.~Park,
  Phys.\ Rev.\ D {\bf 84}, 024044 (2011)  [arXiv:1105.4205 [hep-th]].


\bibitem{Moon:2013jna}
  T.~Moon and Y.~S.~Myung,
  Gen.\ Rel.\ Grav.\  (2013)
  [arXiv:1303.5893 [hep-th]].


\bibitem{Hartnoll:2009sz}
  S.~A.~Hartnoll,
   Class.\ Quant.\ Grav.\  {\bf 26}, 224002 (2009)  [arXiv:0903.3246 [hep-th]].

\bibitem{Hartmann:2013nla}
  B.~Hartmann,
   arXiv:1310.0300 [gr-qc].

\bibitem{Gullu:2009vy}
  I.~Gullu and B.~Tekin,
  Phys.\ Rev.\ D {\bf 80}, 064033 (2009)  [arXiv:0906.0102 [hep-th]].

\end{thebibliography}
\end{document}